\documentclass[preprintnumbers,amsmath,amssymb,floatfix,11pt,prd,onecolumn,superscriptaddress,nofootinbib]{revtex4}
\usepackage{graphicx}
\usepackage{epsfig}
\usepackage{bm}
\usepackage{amsfonts}

\begin{document}

\title{Reconstruction of $f(T)$ and $f(R)$ gravity according to $(m,n)$-type holographic dark energy}

\author{\textbf{M. Umar Farooq}}\email{m_ufarooq@yahoo.com}
 \affiliation{Department of Basic Sciences and Humanities, NUST College of Electrical and Mechanical Engineering,
  Peshawar Road, Rawalpindi, Pakistan}
\author{\textbf{ Mubasher Jamil}} \email{mjamil@camp.nust.edu.pk}
\affiliation{Center for Advanced Mathematics and Physics (CAMP),\\
National University of Sciences and Technology (NUST), H-12,
Islamabad, Pakistan}
\author{\textbf{ Davood  Momeni}}
\email{d.momeni@yahoo.com}
 \affiliation{Eurasian International Center
for Theoretical Physics, Eurasian National University, Astana
010008, Kazakhstan}
\author{\textbf{ Ratbay Myrzakulov}}
\email{rmyrzakulov@gmail.com}\affiliation{Eurasian International
Center for Theoretical Physics, Eurasian National University, Astana
010008, Kazakhstan}

\begin{abstract}
\textbf{Abstract:} Motivated by earlier works on reconstruction of modified gravity models with dark energy components, we extend them by
considering a newly proposed model of ($m,n$)-type of holographic
dark energy for two models of modified gravity, $f(R)$ and $f(T)$ theories, where $R$ and $T$
represent Ricci scalar and torsion scalar respectively. Specifically
we reconstruct the two later gravity models and discuss their
viability and cosmography. The obtained gravity models are ghost  free
, comapatible with local solar system tests and describe
effective positive gravitational constant.

\end{abstract} \maketitle

\newpage
\section{Introduction}
From observational data and different types of cosmological evidences we know that our universe is in acceleration expansion era due to an unknown reason
\cite{ries}. Different kinds of  theories have been introduced  to emplain and clarify this
accelerated behavior of the universe. In the frame of classical 
theory of gravity based on Einstein equations, we label this acceleration to the dark
energy with an unknown essence with a simple candidate as cosmological constant
or vacuum energy of quantum fields spread in the whole space time \cite{cohen}, with two siginificant features as
fine-tuning  of different orders of vacuum energy and cosmic coincidence (for a review see \cite{sami}).

Inspired from string theory , holographic dark energy (HDE) scenario has been constructed  to explain problems of
cosmological constant and others \cite{cohen}. In this scenario we supposed that the total energy of all matter fields in a system with  physical size $L$ must be less than 
the total energy of   a black hole with the same Schwarzschild radius, namely
\begin{equation}
L^3\rho_{V}\leq L M_{p}^2,
\end{equation}
here $\rho_{V}$ denotes the holographic dark energy density as a function of 
the $UV$ cut-off, $L$ as prefered  IR cut-off and $M_{p}$  Planck mass. By saturating the above inequality (1.1) by
choosing the largest length $L$ one identifies the holographic dark
energy:
\begin{equation}
\rho_{V}=3c^2M_{p}^2L^{-2},
\end{equation}
where $c$ is some constant \footnote{Here $c$ is not speed of
light}.

Since the model of holographic dark energy model emerges from one of
the fundamental principles of physics, it has been extremely
successful in explaining numerous cosmological puzzles and has been
a leading candidate of cosmic acceleration problem. Specifically it
can resolve the coincidence problem \cite{pavon} and phantom
crossing \cite{phantom}. Besides the model has reasonable agreement
with the astrophysical data of CMB, SNe Ia and galaxy redshift
surveys \cite{obs}. On these accounts, the HDE paradigm has been
extended via different cut-offs \cite{cut} and entropy corrections
\cite{entropy}. A recent extension of this idea is ($m,n$)-type
holographic dark energy \cite{ling2}, where $m$ and $n$ are the
parameters associated with the chosen IR cut-off at a
phenomenological level (explicitly written in later sections). It is
a generalization of the old models of HDE and agegraphic dark
energy. Reconstruction of modified gravity theories according to HDE
has been discussed in literature before. Independently Setare and Wu
\& Zu \cite{setare} studied correspondence of HDE and $f(R)$
gravity. Recently it has been  investigated  the relations
between HDE and $f(T)$ gravity \cite{d} . 
%These studies have also been
%extended via entropy corrections \cite{karami1}.
 In this article, we
investigate the correspondence of $(m,n$)-type HDE in first modification of Eintein gravity in $f(R)$ and secondly in a recently proposed Weitzenbock model based on scalar torsion
$f(T)$ gravity .
The plan of this paper is as follows: In Sec-II, we study
correspondence of $(m,n$)-type HDE in $f(T)$ gravity. In Sec-III, we
obtain the reconstructed form of $f(T)$ under the assumption of
power-law acceleration. In Sec-IV, we discuss the cosmography for
the $f(T)$ model. In Sec-V, we study correspondence of $(m,n$)-type
HDE in $f(R)$ gravity and discuss its viability in Sec-VI. We
briefly discuss conclusions in Sec-VII. We adopt the natural
gravitational units $c=8\pi G=1.$

\section{$f(T)$ gravity according to the (m,n)-type holographic
dark energy}

In this section we summarrize basics of generalized  teleparallel gravity (GTEGR) of namely $f(T)$ in which we extend the TEGR 
Lagrangian $T$ (which is dynamically equivalent to GR)  to a generic  function $f(T)=T+g(T)$ as we did for 
 the generalization of GR to the modified
$f(R)$ gravity. One form of  the action of $f(T)$ gravity coupled minimally to  matter
$\mathcal{L}_{m}$ is given by \cite{ben}
\begin{equation}
\mathcal{S}=\frac{1}{2}\int d^4xe[T+g(T)+\mathcal{L}_{m}],
\end{equation}
where $e=det(e^i_{\mu})=\sqrt{-g}$ and $\mathcal{L}_{m}$ is the
matter fields Lagrangian density. The TEGR Lagrangian $\mathcal{L}_{TEGR}=T$,is
just the torsion scalar and can be wriiten explicitly as the following:
\begin{equation}
T=S_{\rho}^{\mu\nu}T_{\mu\nu}^{\rho},
\end{equation}
where
\begin{equation}\begin{split}
T_{\mu\nu}^{\rho}&=e_{i}^{\rho}(\partial_{\mu}e_{\nu}^i-\partial_{\nu}e_{\mu}^i,\\
S_{\rho}^{\mu\nu}&=\frac{1}{2}(K_{\rho}^{\mu\nu}+\delta_{\rho}^{\mu}T_{\theta}^{\theta\nu},
-\delta_{\rho}^{\nu}T_{\theta}^{\theta\mu}),
\end{split}\end{equation}
where $K_{\rho}^{\mu\nu}$ is the contorsion tensor
\begin{equation}
K_{\rho}^{\mu\nu}=-\frac{1}{2}(T^{\mu\nu}_{\rho}-T^{\nu\mu}_{\rho}-T_{\rho}^{\mu\nu}).
\end{equation}
Here  dynamical  variable is vierbein tetrads $e_{\mu}^i$, consequently the
field equations are derived  by $\frac{\delta S}{\delta e_{\mu}^i}=0$:
\begin{equation}
e^{-1}\partial_{\mu}(eS_i^{\mu\nu})(1+g_{T})-e_i^{\lambda}T_{\mu\lambda}^\rho
S_{\rho}^{\nu\mu}g_T+S_i^{\mu\nu}\partial_{\mu}(T)g_{TT}-\frac{1}{4}e_i^{\nu}(1+g(T))=
\frac{1}{2}e_i^{\rho}\tau_{\rho}^{\nu},
\end{equation}
where $g_T=\frac{dg}{dT},g_{TT}=\frac{d^2g}{dT^2}$, $\tau_{\rho\nu}$ is the stress tensor. Different aspects of this torsion theory have been investigate \cite{ft}.
We consider cosmology of a spatially-flat $k=0$ universe obeys the modified  Friedmann-Robertson-Walker
(FRW) equations from the metric as:
\begin{equation}
ds^2=dt^2-a(t)^2\sum_{i=1}^{3}(dx^i)^2,
\end{equation}
where $a(t)$ stands for the scale factor and $t$ is the cosmic time.
Because of isotropicity and homogenious assumptions , we assume that the  spacetime contains
perfect fluid matter fields , so  the field equations (2.5), we can write
\begin{equation}\begin{split}
T&=-6H^2\label{t},\\
3H^2&=\rho-\frac{1}{2}g-6H^2g_T,\\
-3H^2-2\dot{H}&=p+\frac{1}{2}g+2(3H^2+\dot{H})g_T-24\dot{H}H^2g_{TT},
\end{split}\end{equation}
here $\rho$ and $p$ are the energy density and pressure of extra fields, $H=\frac{\dot{a}}{a}$ is the Hubble
parameter.

We now propose the correspondence between the $(m,n)-$ type
holographic dark energy scenario and GTEGR $f(T)$ dark energy model. For
$(m,n)$ holographic dark energy density $\rho_{V}$ we write:
\begin{equation}
\rho_{V}=\frac{3b^2}{L^2}\label{rhov},
\end{equation}
with $b=\text{constant}$ and $L=\text{generalized IR cut-off}$ is defined by \cite{ling2}
\begin{equation}
L=\frac{1}{a^{m}(t)}\int_{0}^{t}a^{n}(t')dt',
\end{equation}
which, on using an appropriate transformation can be written as
\begin{equation}
L=\frac{1}{a^{m}(t)}\int_{0}^{t}\frac {a^{n-1}}{H}da.\label{l}
\end{equation}
This model proposed in the level of  a phenomenological model for dark enegy. It is straighforward to show that the model mimics a kind of 
  generalized  agegraphic dark energy  models. But it is an essential difference between the agegraphic model of dark energy and this new proposed alternative. 
The naive difference backs to the choice of the appropriate cut-off of the model. The cut-off is a physically meaningfull and reachable scale of length. Here in this new proposed model as $(m,n)$ type, we select"' the conformal-like age as the holographic
characteristic size"'. Another adventages of this model is  that although the pair of the $(m,n)$ are real arbitary constants, but for some reasonable
 values of $(m, n) $ this extended  model of holographic dark energy suddently  passes the phantom 
cross line $\omega = −1$ . The main motivation of this phantom cross line is that, in spite of the usual scalar models of phantom dark energy, here  there is no need to introducing an interaction function phenomenologically  between  a two component mix fluids of the dark
energy and  matter. This model also is applicable to introduce a new generalized future event horizon as the characteristic size of the model and consequently a new cut-off. We indicate here that in this model the pair of $(m,n)$ are arbitarary and at  level of phenomenologicalthey high energy models, need not be
integers. In general about these parameters we can say,
for age-like holographic models,in the case of $m = n$ it seems that
dark energy has the same time evolution as the dominant ingredient in the early epoches of the
universe, implying that dark energy might be uniﬁed with dark matter, analogous to what
happened in cosmological models with generalized Chaplygin gas. Also we mention here that particle horizon as the holographic characteristic scale in thios model corresponds to 
$m = n = −1$.
Further we want to specify the values of $m$ as a special case. It is easy to show that 
for  $m > 0$, then the HDE is equivalent to a phantom ﬁeld. When $m = 0$, then the holographic dark energy  is just the cosmological constant. Also.
 the new agegraphic dark energy model corresponds to $(n = −1, m = 0)$. Finally if  $−1 < m < 0$, the HDE can drive the universe into an accelerating phase . Clearly we must fix this choice in favor of
observational data . The pairs $(m, n) = (0, −1)$ and $(m, n) = (4,3)$,
are compatiable  with  observation data .

Using the definition of critical energy density $\rho_{cr}=3H^2$, we
can write the dimensionless dark energy parameter as
\begin{equation}
\Omega_{V}=\frac{\rho_{V}}{\rho_{cr}}=\frac{b^2}{H^2L^2}\label{omegav}.
\end{equation}
Apply the definition of $\Omega_{V}$ and $\rho_{cr}$, we can write
\begin{equation}
\dot{L}=\dot{R}_{h}=-mHL+a^{n-m}(t),
\end{equation}
\begin{equation}
\dot{L}=\dot{R}_{h}=a^{n-m}(t)-\frac{mb}{\sqrt{\Omega_{V}}}.
\end{equation}
In dark energy  dominanted
era  it evolves according to the conservation
equation
\begin{equation}
\dot{\rho_{V}}+3H(\rho_{V}+p_{V})=0.
\end{equation}
Differentiating (2.8) and using (2.13), we can write
\begin{equation}
\dot{\rho_{V}}=-\frac{2}{L}\left[a^{n-m}(t)-\frac{mb}{\sqrt{\Omega_{V}}}\right]\rho_{V}.
\end{equation}
So with this, Eq. (2.14) becomes
\begin{equation}
-\frac{2}{L}\left[a^{n-m}-\frac{mb}{\sqrt{\Omega_{V}}}\right]+3H(1+\omega_{V})=0,
\end{equation}
and we can obtain
\begin{equation}
\omega_{V}=\frac{2}{3}\frac{\sqrt{\omega_{V}}}{b}a^{n-m}(t)-\frac{2m+3}{3}
\end{equation}
Now  equations (2.7) can be rewritten:
\begin{equation}
3H^2=\rho+\rho_{V},\hspace{1.3cm}\rho_{V}=-\frac{1}{2}g-6H^2g_{T},
\end{equation}
\begin{equation}
-3H^2-2\dot{H}=p+p_{V},\hspace{1.2cm}p_{V}=\frac{1}{2}g+2(3H^2+\dot{H})g_{T}-24\dot{H}H^2g_{TT}.
\end{equation}
Combining above equations (2.18) and (2.19) we get
\begin{equation}
\rho_{V}+p_{V}=2\dot{H}g_{T}-24\dot{H}H^2g_{TT}.\label{eq}.
\end{equation}
While on using $p_{V}=\omega_{V}\rho_{V}$ and the values of $H$ and
$\dot{H}$ we can find the general form of the $g(T)$ from the
(\ref{eq}).

\section{Power law acceleration}

We assume that the accelerated expansion of the whole universe is
governed by a power law scale factor, like
\begin{equation}
a(t)\sim t^p,\ \ p>1
\end{equation}
Using this simple but physically reasonable  assumption, we have the following quantities from
 (\ref{l}),(\ref{omegav}),(\ref{t}),(\ref{rhov})
\begin{eqnarray}
L&=&\frac{t^{1+p(n-m)}}{pn+1},\\
\Omega_V&=&\Big[\frac{b(pn+1)}{p}\Big]^2 t^{2p(m-n)},\\
T&=&-\frac{6p^2}{t^2},\\
\rho_V&=&3\Big[b(pn+1)\Big]^2\Big[-\frac{6p^2}{T}\Big]^{-1+p(m-n)}.
\end{eqnarray}
Substituting these expressions in (\ref{eq}) and by a change of the variable to the $x=-T$, we obtain
\begin{equation}
g_{xx}+\frac{g_x}{2x}=-\frac{A}{2}x^{-1-p(m-n)},
\end{equation}
which gives us the following elementary solution
\begin{eqnarray}
g(T)=\frac{A}{(1-p(m-n))(2p(m-n)-1)}(-T)^{1-p(m-n)}+2c_1\sqrt{-T}+c_2,\label{model}
\end{eqnarray}
where
$$A=-2\sqrt{p}\rho_V^0\Big[-2+2p(m-n)\Big](6p^2)^{-\frac{3}{2}+p(m-n)}.$$
This GTEGR model is a natural extension of our former model of generalized
teleparallel gravity \cite{attractor}. If $m=n$, then this model
coincides with \cite{attractor}.
% The first and the third terms
%(excluding the middle term) have correspondence with the
%cosmological constant EoS in torsion gravity. There are many kinds
%of such models, reconstructed from different kinds of dark energy
%models. For example the form  may be inspired from a model for dark
%energy from a proposed form of the Veneziano ghost \cite{karami}.
%But the linear term is needed to show the differences between our
%results in $f (T )$ gravity from the Einstein gravity.
 From the
cosmography \cite{cosmography} and using data of
BAO, Supernovae Ia and WMAP, we notice that in this case with $m=n$ which is equivalence to the extended particle horizon as cut-off,
if we choose
$$
A=\Omega_{m0},\ \ c_1=\sqrt{6}H_0(\Omega_{m0}-1),\ \ c_2=0,
$$
%then we can estimate
%the parameters of our proposed $f (T )$ model as a function of
%Hubble parameter $H_0$ and the cosmographic parameters and
%the value of matter density parameter.

\section{Cosmography}
But for general model, proposed in (\ref{model}) it is needed to check that the cosmography parameters. We need that
\begin{eqnarray}
g(T_0)=6H_0^2(\Omega_{m0}-1),\ \ g'(T_0)=0,\\6H_0^2(1+g''(T_0))=\frac{1}{2}-\frac{3\Omega_{m0}}{4(1+q_0)}  \label{test}
\end{eqnarray}
It's better we write the (\ref{model}) in the following form
\begin{eqnarray}
g(T)=\alpha(-T)^{\mu}+2c_1\sqrt{-T}+c_2,\label{model2}
\end{eqnarray}
where $$ \alpha=\frac{A}{(1-p(m-n))(2p(m-n)-1)},\ \ \mu=1-p(m-n).$$
By substituting (\ref{model2}) in (\ref{test}) we obtain
\begin{equation}
\alpha=-3\,{\frac {{H_{{0}}}^{2} \left( 24\,{H_{{0}}}^{2}+24\,{H_{{0}}
}^{2}q_{{0}}+3\,\Omega_{{{\it m0}}}-2-2\,q_{{0}} \right) }{{6}^{\mu}{H
_{{0}}}^{2\,\mu}\mu\, \left( 2\,\mu\,q_{{0}}+2\,\mu-1-q_{{0}} \right)
}}
\end{equation}

\begin{equation}
c_{{1}}=\frac{1}{2}\,{\frac {H_{{0}} \left( 24\,{H_{{0}}}^{2}+24\,{H_{{0}}}^{2
}q_{{0}}+3\,\Omega_{{{\it m0}}}-2-2\,q_{{0}} \right) \sqrt {6}}{2\,\mu
\,q_{{0}}+2\,\mu-1-q_{{0}}}}
\end{equation}

\begin{equation}
c_{{2}}=-{\frac {\Gamma }{\mu\, \left( 2\,\mu\,q_{{0}}+2\,\mu-1-q_{{0}
} \right) }},
\end{equation}
where
\begin{eqnarray}
\Gamma =3\,{H_{{0}}}^{2} ( 4\,{\mu}^{2}q_{{0}}+4\,{\mu}^{2}-2\,
\mu-2\,\mu\,q_{{0}}-4\,\Omega_{{{\it m0}}}{\mu}^{2}q_{{0}}\\\nonumber-4\,\Omega_{
{{\it m0}}}{\mu}^{2}+2\,\Omega_{{{\it m0}}}\mu+2\,\Omega_{{{\it m0}}}
\mu\,q_{{0}}+6\,\mu\,\Omega_{{{\it m0}}})\\ \nonumber+
3\,{H_{{0}}}^{2}(48\,\mu\,{H_{{0}}}^{2}q_{{0
}}+48\,\mu\,{H_{{0}}}^{2}-4\,\mu -4\,\mu\,q_{{0}}-24\, \left( -1
 \right) ^{\mu}{H_{{0}}}^{2}\\ \nonumber-24\, \left( -1 \right) ^{\mu}{H_{{0}}}^{2
}q_{{0}}-3\, \left( -1 \right) ^{\mu}\Omega_{{{\it m0}}}+2\, \left( -1
 \right) ^{\mu}+2\, \left( -1 \right) ^{\mu}q_{{0}} )
\end{eqnarray}

\section{Reconstruction of $f(R)$ gravity according to the (m,n)-type holographic
dark energy}
To reconstruction of FRW cosmology in $f (R)$  (for
general review, see \cite{fr} and \cite{frme}) we start by the following action:
\begin{eqnarray}
S=\int d^4x \sqrt{-g}\Big(\frac{f(R)}{2}+\mathcal{L}_m\Big)
\end{eqnarray}
The  first FRW equation is:
\begin{eqnarray}
-\frac{f}{2}+3(H^2+\dot H)f'-18(4H^2\dot H+H \ddot H)f''+\rho_V=0
\end{eqnarray}
Again by assumning the power law expansion $a(t)\sim t^p,\ \ p>1$  as the previous f(T) reconstruction,using
\begin{eqnarray}
R=\frac{6p(2p-1)}{t^2},\\
\rho_V=3\Big[b(pn+1)\Big]^2t^{2(-1+p(m-n))}.
\end{eqnarray}
So
\begin{eqnarray}
\rho_V=3\Big[b(pn+1)\Big]^2\Big(\frac{6p(2p-1)}{R}\Big)^{2(-1+p(m-n))} .
\end{eqnarray}
Thus
\begin{eqnarray}
-\frac{f}{2}+\Big(\frac{p-1}{2(2p-1)}\Big)Rf'-\frac{1}{1-2p}R^2f''+3
\Big[b(pn+1)\Big]^2\Big(\frac{6p(2p-1)}{R}\Big)^{2(-1+p(m-n))}=0\label{eqfr}.
\end{eqnarray}
The solution for (\ref{eqfr}) reads
\begin{eqnarray}
f \left( R \right) =C_1{R}^{\frac{1}{4}(-p+3-\,\sqrt {{p}^{2}+
26\,p-7})}{\it }+C_2{R}^{\frac{1}{4}(-p+3+\,\sqrt {{p}^{2}+26\,p-7})}{
\it }\\ \nonumber-
\,{\frac { 72\left( pn+1 \right) ^{2}\left( p-1/2 \right) ^{2}{\kappa}^{2}p}{2+4\, \left( m-n
 \right)  \left( m-1/4-n \right) {p}^{2}+ \left( -1-5\,m+5\,n \right){b}^{2}
p}} {R}^{2+ \left( -2\,m+2\,n \right) p
}\label{solfr}.
\end{eqnarray}
To recovering the usual HDE model with particle horizon we consider  the case $m=n=1$, we have
\begin{eqnarray}
f \left( R \right) =C_1{R}^{\frac{1}{4}(-p+3-\,\sqrt {{p}^{2}+26\,p-7})}{\it }+C_2{R}^{\frac{1}{4}(-p+3+\,\sqrt {{p}^{2}+26\,p-7})}{
\it }-\,{\frac { 72\left( p+1 \right) ^{2}{b}^{2} \left( p-\frac{1}{2}
 \right) ^{2}p}{2-p}}{\kappa}^{2}{R}^{2}
\end{eqnarray}

\section{Viability conditions of  (m,n)-type holographic
dark energy- $f(R)$ theories}

Based on cosmological viable $f(R)$ models tests in Ref. \cite{segei2006} the essential constraints  and
limitations \cite{Pogosian} for action:
\begin{equation}
S_{g}=\frac{1}{16\pi G}\int d^{4}x\sqrt{-g} f(R)  \nonumber
\end{equation}
to be consistent with gravitational (Solar system tests) and cosmological (Large scale behaviors)
are:
\begin{enumerate}
\item $f^{\prime \prime }(R)\geq 0$ for $R\gg f^{\prime \prime }(R)$\cite{Faraoni} .
\item $1+f^{\prime }(R)>0$ to avoiding of ghost \cite{Nunez}.
\item $f^{\prime }(R)<0$.
\item $f^{\prime }(R)$ must be very small in recent cosmological dark energy dominant era\cite{Sawicki}.
\end{enumerate}

To examine these  conditions (1-4) for a general $f(R)$
model described in (\ref{solfr}) we observe that  validity  conditions (1-4) depends on the parameters  $m$ and
$n$. For generality, we consider the only case $m\neq n$.

\begin{enumerate}
\item $f^{\prime \prime }(R)=R^{\delta_{-}-2}$ for $R\gg R_0$. Thus this model
is stable for ultra voiolent (UV) completion of GR with an effective
gravitational constant  $G_\text{eff}\equiv \frac{G}{1+c
R^{\delta_{-}-2}},\ \ \delta_{-}>1$ . Obviously in the very  early universe  the curvature was very high
when $R>>R_0\Longrightarrow G_\text{eff}\simeq G,\ \ -1<\delta_{-}<0$.

\item To avoiding the guest problem in UV regime we must check  $R\gg R_0$,$1+f^{\prime }(R)>0$ . It means  $
G_\text{eff}>0$ so theory is ghost free due to effects of modified gravity.

\item $\frac{f(R)}{R}|_{R\rightarrow\infty}\rightarrow\infty$ . Further  $f^{\prime }(R)\rightarrow\infty$. So the  model recovers GR in the begining UV regime.

\item For our model in UV regime  $f^{\prime }(R)<<1,\ \ R$. So this model confirms  the local
solar system tests. Also we estimate it numerically to find:
 $$|R^{\delta_{-}-2}|<10^{-6}\Longrightarrow |\delta_{-} |<\frac{\log (10^{-6})}{\log R}+2,R\approx O(\Lambda)$$.
\end{enumerate}
So our f(R) model reconstructed from HDE is a ghost free and observationally viable according to viability conditions.

\begin{figure}
\centering
 \includegraphics [scale=0.4]{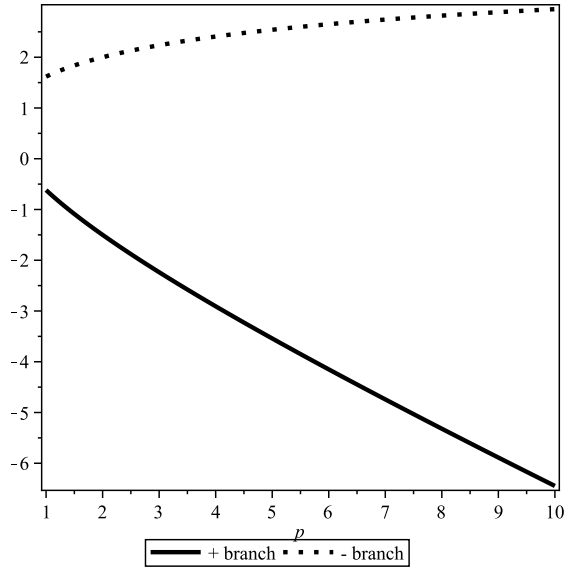}% scale goes from 0 to 1.
  \caption{Variation of the $\delta_{\pm}=\frac{1}{4}(-p+3\pm\,\sqrt {{p}^{2}+26\,p-7})$ for $p>1$. Note that $\delta_{-}>1,\ \ \delta_{+}<-\frac{1}{2}$. So at late time , i.e. at $R>>R_0$, the power $\delta_{-}$ is dominant.}
  \label{1.eps}
\end{figure}

\section{Conclusion}

In this paper, we studied a correspondence between a newly proposed
model of $(m,n)-$ type HDE model with two modified theories of
gravity namely $f(R)$ and $f(T)$ gravity. This study generalizes
some previous studies such as the papers cited in the abstracts. For
our obtained $f(T)$ model, we calculated the cosmographic parameters
while for the calculated $f(R)$ model, we investigated the stability
and viability conditions. It is shown that our model of $f(R)$
recovers to general relativistic limit at early times, free from
ghosts, and consistent with the solar system tests. So, we reconstructed a newly motivated model of holographic dark energy in the frame work of models of modified gravity, one in usual Riemannian form f(R) and another with Weitzenbock connections in the frame work of modified Einstein-Cartan theory.

\subsection*{Acknowledgments}
The work of M. Jamil has been  supported from the research grant
provided by the Higher Education Commission, Islamabad, Pakistan.

\end{document}